\documentclass[aps,prl,twocolumn,preprintnumbers,amsmath,amssymb]{revtex4}
\usepackage{amsmath}
\usepackage{amssymb}
\usepackage{bm}
\usepackage{graphicx}
\usepackage{eucal}
\usepackage{mathrsfs}
\usepackage{theorem}
\usepackage{xcolor}
\usepackage{multirow}
\usepackage{bigdelim}
\usepackage{cancel}
\usepackage[normalem]{ulem}
\usepackage{graphicx}
\usepackage[caption=false]{subfig}
\usepackage[unicode=true,pdfusetitle,bookmarks=true,
bookmarksnumbered=false,bookmarksopen=false,breaklinks=false,
pdfborder={0 0 0},backref=false,colorlinks=true,citecolor=red]
{hyperref}
\usepackage{array}
\newcommand{\dd}{\mathrm{d}}

\newcommand{\del}{\nabla}
\renewcommand{\b}[1]{{\bf #1}}

\newcommand{\beq}{\begin{equation}}
\newcommand{\eeq}{\end{equation}}

\newcommand{\Kim}{K_\text{Im}}

\begin{document}

\title{Kirigami mechanics as stress relief by elastic charges}


\author{Michael Moshe$^{a,b}$}
\email[]{mmoshe@syr.edu}
\author{Edward Esposito$^{c}$}
\author{Suraj Shankar$^{b,d}$}
\email[]{sushanka@syr.edu}
\author{Baris Bircan$^{e}$}
\author{Itai Cohen$^{c}$}
\email[]{itai.cohen@cornell.edu}
\author{David R. Nelson$^{a}$}
\email[]{nelson@physics.harvard.edu}
\author{Mark J. Bowick$^{b,d}$}
\email[]{bowick@kitp.ucsb.edu}

\affiliation{$^a$Department of Physics, Harvard University, Cambridge, Massachusetts 02138, USA.\\
$^b$Physics Department and Syracuse Soft and Living Matter Program, Syracuse University, Syracuse, NY 13244, USA.\\
$^c$Laboratory of Atomic and Solid State Physics, Cornell University, Ithaca, NY 14853, USA.\\
$^d$Kavli Institute for Theoretical Physics, University of California, Santa Barbara, CA 93106, USA.\\
$^e$School of Applied and Engineering Physics, Cornell University, Ithaca, NY 14853, USA.}



\begin{abstract}
	We develop a geometric approach to understand the mechanics of perforated thin elastic sheets, using the method of strain-dependent image elastic charges. This technique recognizes the buckling response of a hole under external load as a geometrically tuned mechanism of stress relief. We use a diagonally pulled square paper frame as a model system to quantitatively test and validate our approach. Specifically, we compare nonlinear force-extension curves and global displacement fields in theory and experiment. We find a strong softening of the force response accompanied by curvature localization at the inner corners of the buckled frame. Counterintuitively, though in complete agreement with our theory, for a range of intermediate hole sizes, wider frames are found to buckle more easily than narrower ones. Upon extending these ideas to many holes, we demonstrate that interacting elastic image charges can provide a useful \emph{kirigami} design principle to selectively relax stresses in elastic materials.
\end{abstract}


\maketitle


Kirigami, the art of cutting and folding paper has emerged as a powerful tool to dramatically modify, reconfigure and program material properties \cite{mullin2007pattern,Bertoldi2010,florijn2014programmable,Song2015,Zhang2015,Lamoureux2015,Shyu2015,Wu2016,dias2017kirigami,Rafsanjani2017}. Since kirigami is scale invariant, it can be combined with rapid miniaturization to design metamaterial response and structures at the smallest scales \cite{rogers2009curvy,deng2016wrinkled}. Such approaches were recently demonstrated in graphene \cite{Blees2015} and now provide unprecendented opportunities for designing devices with novel electronic and mechanical properties.  With the advent of such technologies, it has become increasingly important to characterize and understand the various ways in which material deformations accomodate stress and stress relaxation through instabilities in thin two dimensional ($2d$) elastic sheets \cite{eran2004leaves,genzer2006soft,witten2007stress,li2012mechanics,reis2015designer,bertoldi2017flexible,dias2017kirigami}.

\begin{table}[t]
	\centering
	{
		{\setlength{\extrarowheight}{4pt}
		\begin{tabular}{c c l}
			\hline
			\hline
			\hspace{0.5em}Configuration\hspace{1em} & Aspect ratio\hspace{1em} & \multicolumn{1}{c}{$k_{\mathrm{eff}}$}\\[4pt]
			\hline\\[-8pt]	
			Planar & -- & $\propto Y\left[\Phi\left(w/L\right)\right]^2\left(\dfrac{w}{L}\right)^2$\\[8pt]
			\multirow{3}{*}{ Buckled }
			\ldelim\{{3}{3mm} & $\dfrac{1}{8}<\dfrac{w}{L}<\dfrac{1}{4}$ & $\propto\dfrac{\kappa}{L^2}\left[\Phi\left(w/L\right)\right]^2\ln\left(\dfrac{w}{a}\right)$\\[8pt]
			& $\dfrac{w}{L}<\dfrac{1}{8}$ & $\propto\dfrac{\kappa}{L^2}\left(\dfrac{w}{L}\right)$\\[8pt]
			\hline
			\hline
	\end{tabular}}
	}
	\caption{A summary of the effective spring constants $k_{\mathrm{eff}}$ for different frame aspect ratios $w/L$, for planar and buckled configurations.}
	\label{table:keff}
\end{table}

The mechanics of thin elastic sheets is controlled by the dimensionless F{\"o}ppl-von K{\'a}rm{\'a}n (FvK) number $\gamma=YR^2/\kappa$ \cite{LandauElasticityBook} that indicates the relative ease of in-plane stretching versus out-of-plane bending. Here $R$ is a characteristic linear dimension of the sheet, $Y$ the $2d$ Young's modulus and $\kappa$ the bending rigidity. Under external load, a thin sheet trades energetically expensive stretching with bending to relieve stresses by either buckling \cite{LandauElasticityBook} or wrinkling \cite{Cerda2002,Cerda2003,King12}, possibly followed by secondary instabilities \cite{benamar97,pocivavsek2008stress,witten2007stress}.
By introducing holes or cuts, kirigami now provides a distinct route to locally relieve stresses through these geometric features, though a general characterization of its effective mechanical response is not known.
The \emph{inverse problem} of predicting the correct kirigami pattern to relax a given pre-stress in a material also remains an open problem, complicated by the notorious nonlinearity inherent in thin sheet elasticity.

\begin{figure}
	\includegraphics[width=0.47\textwidth]{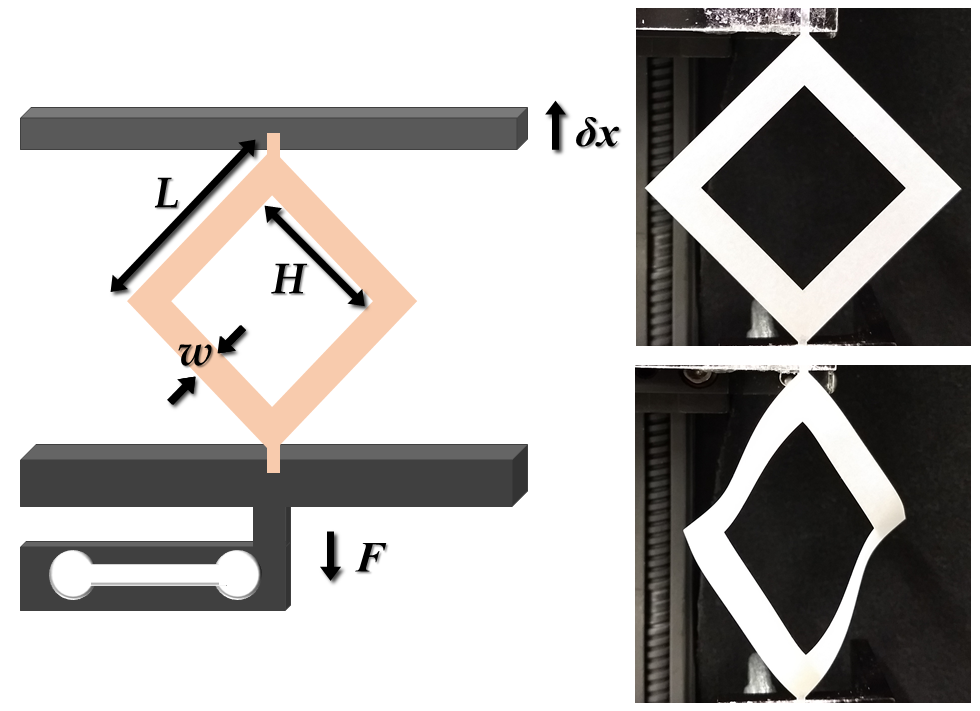}
	\centering
	\caption{{\bf Force displacement measurement}. (a) Square paper frames with thin tabs on two opposite corners are glued to two opposed aluminum plates. The top plate can be displaced upward using a micrometer screw-gauge for very fine displacements, or using a stepper motor for larger displacements. A load cell attached to the bottom plate measures the force required to hold the frame at fixed displacement. In our experiments, all frames had the same side length $L$ of $5.04$~cm, while the frame width $w = (L-H)/2$ was varied. (b) Paper frame in a planar configuration. (c) Paper frame in a buckled configuration.
	}
	\label{fig:Illustration}
\end{figure}

In this paper, we develop a geometric framework to address some of the general mechanical consequences of kirigami and report on experimental measurements of force-extension curves of pulled paper frames. A more detailed theoretical and numerical analysis is presented in a longer companion paper \cite{Theorypaper}. Starting with a single square frame, we use the technique of strain-dependent image elastic charges to show that a hole under external load acts as a geometrically tunable source of local stress, which is relaxed by local buckling. The lowest order image elastic charge induced in a hole is a quadrupolar singularity in Gaussian curvature.  When permitted by the shape of the hole, this singularity can fractionalize into partial disclinations, naturally explaining the curvature localization at interior corners seen experimentally for square frames. Thus, the buckling response of the sheet can be viewed as the sheet screening the image charges by adopting a curved $3d$ configuration that leads to a softer force response. Through a quantitative comparison between theory and experiment, we confirm predictions for the geometric dependence of effective spring constants (summarized in Table~\ref{table:keff}) and the buckling threshold, along with more local measures of deformation like the full displacement field.

Similar buckling induced softened mechanical response has been previously investigated in periodic arrays of slits under uniaxial tension \cite{Shyu2015,Midori2016,Rafsanjani2017}. Our framework rationalizes these previous results, extends to arbitrary hole shapes \footnote{Much of our focus is on square holes, because the concentration of elastic charges at sharp corners simplifies the analysis; elastic charges are delocalized around the rim when one considers, say, circular holes instead.}, and provides a systematic approach to handle many holes. In addition, collective effects arising from interactions between holes are neglected in works that just analyze the unit cell of a periodic lattice, but are easily captured using the elastic charge framework. Using a flattened cone as an example, we demonstrate how interactions between image charges can guide the design of appropriate kirigami patterns to relax the pre-existing stress in the system.

\begin{figure*}
\includegraphics[width=\linewidth]{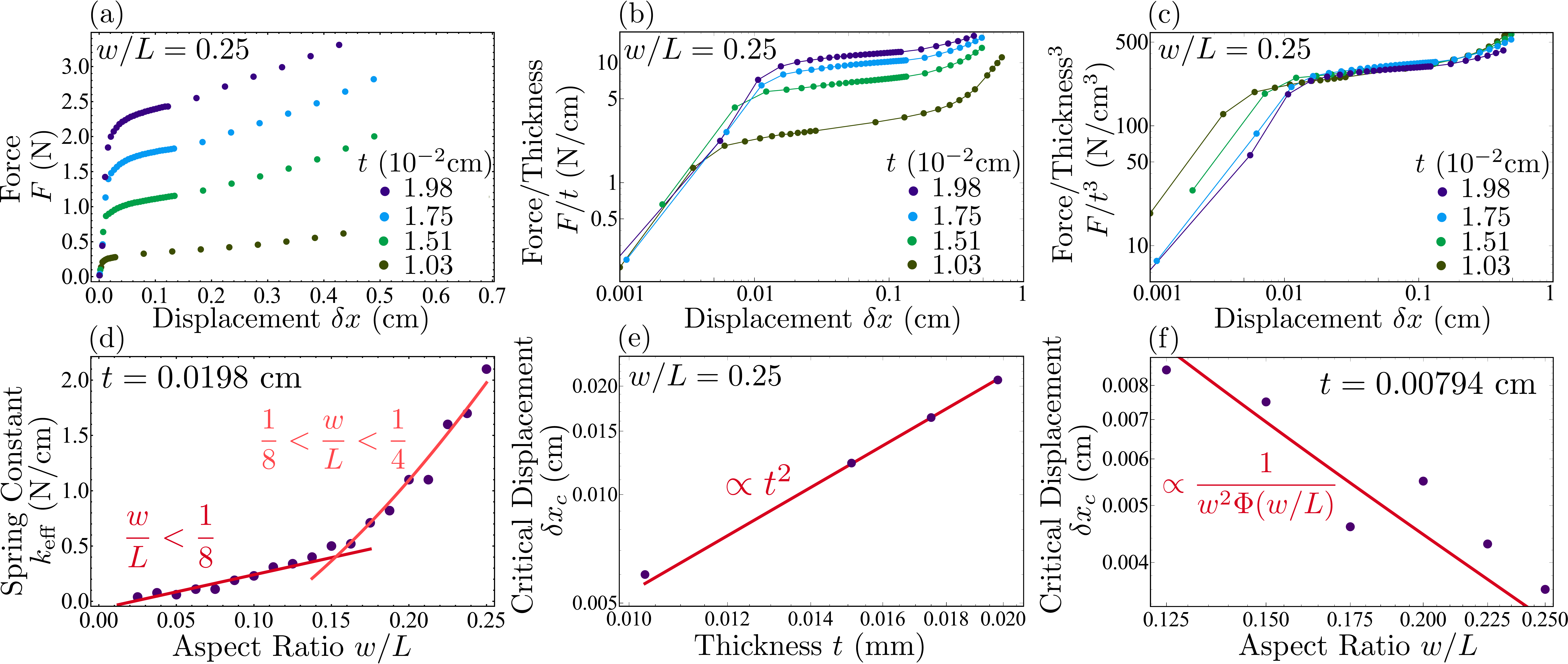}
\centering
\caption{{\bf Experimental measurements of square frames subjected to tensile load along the diagonal.} (a)  Force-displacement curves for frames with $w/L = 0.25$ and thicknesses varying between $0.01$ and $0.02$~cm. (b) When normalized by thickness, curves collapse at small displacement, confirming that the frames are planar at this regime. (c) When normalized by thickness cubed, curves collapse in the post-buckling regime, confirming that energy increase is predominately bending. (d) Effective spring constant in the post-buckled regime as function of frame aspect ratio $w/L$ in the intermediate and large hole regimes for a frame of thickness $t=0.0198$~cm, confirming the multi-scale behavior in (Table~\ref{table:keff}). The curve in the large hole regime is linear \eqref{eq:keff2}, while the curve in the intermediate hole regime corresponds to \eqref{eq:keff1}, with the pre-factor $c$ and regularizing cutoff $a$ taken as fitting parameters. (e) Critical displacement as function of thickness for a frame of $w/L=0.25$, growing as $t^{1.9}$ (solid line), in good agreement with Eq.~\eqref{eq:CritDisp}. (f) Critical displacement as function of the frame's aspect ratio for a frame of thickness $t=0.00794$~cm in the intermediate hole size regime, in agreement with Eq.~\eqref{eq:CritDisp} (solid line).}
\label{fig:ExpData}
\end{figure*}

The square frames we study were cut from sheets of Glama Natural paper of various thicknesses ($t=0.01-0.02$~mm) with an edge length $L=5.04$~cm. The aspect ratio $w/L$, where $w=(L-H)/2$ is the frame width and $H$ is the hole size (see Fig.~\ref{fig:Illustration}), was varied between $0<w/L<1/4$. Frames with larger aspect ratios corresponding to smaller holes often failed before buckling. To measure force-extension curves, we mount the opposite outer corners of the frames between a top and bottom plate and extend them by a distance $\delta x$ (fig. \ref{fig:Illustration}).  To generate displacements we use a rail guided Haydon-Kerk stepper motor to move a micrometer translation stage to which one corner of the frame is attached. Coarse displacements ($\sim500\mu$m) are generated with the stepper motor while finer displacements ($\sim50\mu$m), primarily near buckling, are generated using the micrometer stage. Force measurements were made using a Loadstar parallel cantilever loadcell attached to the bottom plate.  Further details of the experimental protocol are given in the SI. We observe a steep increase in force at low displacements, followed by a leveling off beyond a critical displacement, and finally an increase at high displacements just before the frames tear (Fig.~\ref{fig:ExpData}a).

The initial buckling can be understood within the framework of a stretching to bending transition. The mechanics of the frame is governed by an elastic energy involving both stretching and bending terms quadratic in the stress tensor (${\bm\sigma}$) and the curvature tensor ($\b{b}$). Upon minimizing, we obtain the covariant F{\"o}ppl-von K{\'a}rm{\'a}n (FvK) equations \cite{Koiter1966,ciarlet1997mathematical},
\begin{subequations}
\begin{align}
	\dfrac{1}{Y}\Delta\Delta\chi&=\Kim-K_G\ ,\label{eq:CovFVK1}\\
	\kappa\Delta\mathrm{tr}(\bf{b})&=\sigma^{\mu\nu} b_{\mu\nu}\ .\label{eq:CovFVK2} 
\end{align}
\end{subequations}
Here we have used the $2d$ Airy stress function $\chi$ ($\sigma^{\alpha\beta}=\epsilon^{\alpha\mu}\epsilon^{\beta\nu}\del_{\mu}\del_{\nu}\chi$) and the extrinsic curvature tensor is defined by $b_{\alpha\beta}=\hat{\b{n}}\cdot\del_{\alpha}\b{t}_{\beta}$, where $\hat{\b{n}}$ and $\b{t}_{\beta}$ are the local normal and tangent vectors to the surface, and whose determinant gives the Gaussian curvature $K_G$ of the surface. In terms of a $3d$ Young's modulus $\bar{Y}$, we have $Y=\bar{Y}t$ and $\kappa=\bar{Y}t^3/[12(1-\nu^2)]$, where $t$ is the sheet thickness and $\nu$ is the three dimensional Poisson ratio \cite{AudolyPomeauBook}. Unlike the conventional FvK equations for thin plates, we have additionally included a source of Gaussian curvature $\Kim$ that plays the same role that defects play in crystals \cite{nelson2002defects}, though in our case this function describes a distribution of \emph{image} elastic charges that are induced within the hole and depend on the external load, serving to enforce the appropriate boundary conditions required by the presence of the hole \cite{Theorypaper}. Here the analogy with electrostatics helps, in that the hole under external stress functions like a conductive shell in an external electric field. This framework allows for understanding the various scalings observed in the data.

For very small diagonal displacements ($\delta x<\delta x_c$, with $\delta x_c$ the buckling threshold), it is clear that the frame responds linearly by stretching (Fig.~\ref{fig:ExpData}b). Though the frame is still planar, the effective spring constant is modified by the hole geometry. Setting $\b{b}=\b{0}$ ($K_G=0$), we only have $\Kim$, the image elastic charge, present within the hole. Unlike genuine topological disclinations or dislocations (monopole and dipole singularities) that \emph{cannot} be created by any local deformation \cite{Kupferman2013ARMA}, the leading order contribution to $\Kim$ is a quadrupolar form \cite{Moshe2015PNAS}. In $2d$ the quadrupole can be written as $\b{Q}=Q(\cos2\psi,\sin2\psi)$, $\psi$ being its orientation and $Q$ its magnitude.
In the presence of sharp corners in the hole geometry the induced image elastic charge can fractionalize into \emph{partial} disclinations that localize at the sharp corners, just as in the electrostatic analogue, and generate stress fields similar to their topological counterparts \cite{Seung88}. The partial disclinations have a charge that continuously depends on the external strain imposed, given by $s\equiv Q/H^2=(\delta x/L)\Phi(w/L)$, where $\Phi(w/L)$ is a rational function of the frame's aspect ratio that encodes the hole geometry \cite{Theorypaper}. As $w\to L/2$ (no hole), $\Phi(w/L)\propto(1-2w/L)^2$ vanishes as expected and remains finite in the opposite narrow frame limit ($w\to 0$). This setup allows us to estimate the energy due to stretching and bending.

Prior to buckling, the elastic energy of the planar square frame is approximately $E\sim Ys^2w^2$. For large displacements, the frame buckles allowing $K_G\neq 0$. As the frame's large FvK number ($\gamma=Yw^2/\kappa\gg1$) favors isometric deformations, the frame screens out the induced image charge $\Kim$ with real Gaussian curvature $K_G$ \cite{Seung88}, permitting the stress free state $\chi=0$ to become available. By virtue of the localized partial disclinations, the buckled frame adopts a locally conical shape near the inner corners leading to the energy being $E\sim\kappa[c_1 s+c_2 s^2]\ln(w/a)$, where $a\sim t$ is a microscopic core cut-off and $c_1,c_2$ are numerical constants \cite{Seung88}. Since $F=\dd E/\dd\delta x$, we rescale the force-extension curve by $t$ and $t^3$ for a given aspect ratio ($w/L=0.25$) in Fig.~\ref{fig:ExpData}b,c. We find excellent collapse in the pre-buckling and post-buckling regimes, which are controlled by $Y$ and $\kappa$ respectively. The $t^3$ scaling in the post-buckling plateau indicates the force response is governed by $\kappa$ and the hole geometry alone.

The fractionalized quadrupole naturally delineates different geometric regimes. For $w/L<1/4$, the partial disclinations remain well separated and essentially non-interacting, allowing one to approximately superpose their buckled solutions, while for narrower frames with $w/L<1/8$ \cite{Theorypaper}, higher order charges become important. Within the intermediate frame regime ($1/8<w/L<1/4$), the effective linearized spring constant of the frame post-buckling is then given as
\begin{equation}
	k_{\mathrm{eff}}=\left.\dfrac{\dd^2 E}{\dd\delta x^2}\right|_{0}\propto\dfrac{\kappa}{L^2}\ln\left(\dfrac{w}{a}\right)\left[\Phi\left(w/L\right)\right]^2\ .\label{eq:keff1}
\end{equation}
To calculate the buckled force response of narrow frames ($w/L<1/8$) we need to use an alternate approach.  Here, an infinite series of multipolar charges higher than the quadrupole becomes important, suggesting the appropriate weakly-interacting degrees of freedom are not elastic charges. Instead, we treat the frame edges as quasi-$1d$ ribbons joined in a ring. Neglecting the high energy splay modes, the bending and twisting elastic energy of a ribbon is approximated by $E\sim \kappa w L(\delta\theta/L)^2$, where $\delta\theta\propto\delta x/L$ is the net rotation of the ribbon across its length \cite{AudolyPomeauBook}. Once again computing the effective linearized spring constant for the buckled narrow frames, we obtain
\begin{equation}
	k_{\mathrm{eff}}\propto\dfrac{\kappa w}{L^3}\ .\label{eq:keff2}
\end{equation}
The disparate geometric scaling of $k_\mathrm{eff}$ for different frame widths (Eqs.~\ref{eq:keff1},~\ref{eq:keff2}) is a signature of multi-scale behavior, induced here by the geometry of the hole. The experimentally measured spring constants of the buckled frames agree very well with our theoretical predictions as shown in Fig.~\ref{fig:ExpData}d. The geometric dependence of various linearized spring constants is also summarized for both buckled and planar frames in Table~\ref{table:keff}.

\begin{figure}
	\includegraphics[width=\columnwidth]{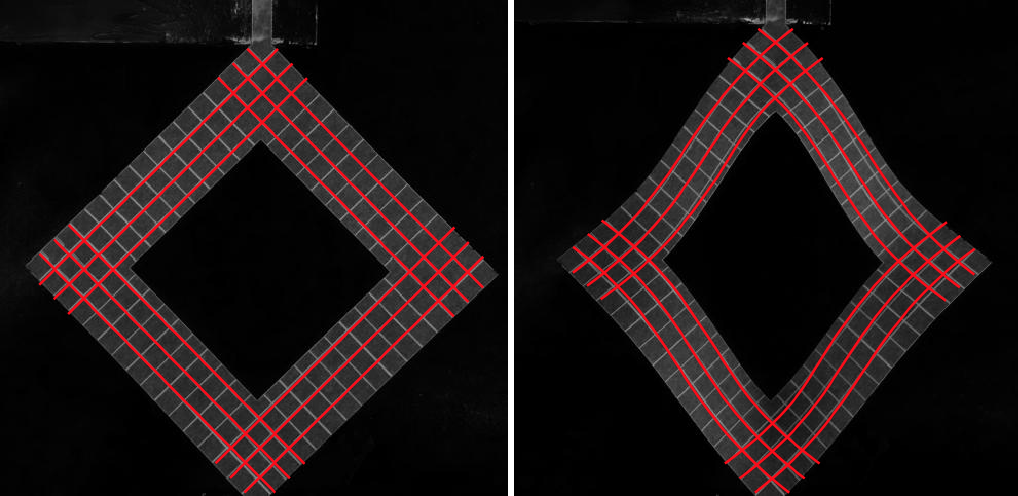}
	\caption{{\bf Comparison between predicted and observed deformations in a buckled frame} (a) An undeformed frame with laser printed Cartesian mesh (gray) and a set of parametric lines (red) fitted to the printed mesh. (b) A deformed frame. Here, the red lines are computed from theory using the original parametric lines as a starting point and fictitious elastic charges as fitting parameters.}
	\label{fig:defmesh}
\end{figure}

For intermediate frame widths $1/8<w/L<1/4$, given that the frame mechanics is dictated by the partial disclinations, we can estimate the geometry dependence of the frame's buckling threshold $\delta x_c$ by adapting previous results on the buckling of topological disclinations \cite{Seung88}. As the region of influence of the partial disclination is a corner plaquette of area $\sim w^2$, using the relevant FvK number $\gamma=Yw^2/\kappa$, we obtain a threshold charge $|s_c|\simeq\gamma_c/\gamma$ ($\gamma_c\approx120$) in order to buckle. Upon using $s=(\delta x/L)\Phi(w/L)$, we find the critical strain,
\begin{equation}
	\dfrac{\delta x_c}{L}\propto\dfrac{1}{\Phi(w/L)}\left(\dfrac{t}{w}\right)^2\ , 
    \label{eq:CritDisp}
\end{equation} 
where we have used the fact that $\kappa/Y\propto t^2$. The quadratic scaling of $\delta x_c$ with $t$ is consistent with observed data (Fig.~\ref{fig:ExpData}e). The dependence of $\delta x_c$ on the frame width $w$ captures crucially the geometric tunability of the local propensity to relax stresses via buckling. Though we expect ultra-narrow frames ($w\to 0$) to have a vanishing threshold for buckling due to sheer loss of material, within the intermediate range of hole sizes Eq.~\ref{eq:CritDisp} in fact suggests a counter-intuitive trend, with wider frames buckling prior to narrow ones. This feature is observed for a thin enough sheet in Fig.~\ref{fig:ExpData}f.

\begin{figure}
	\includegraphics[width=\columnwidth]{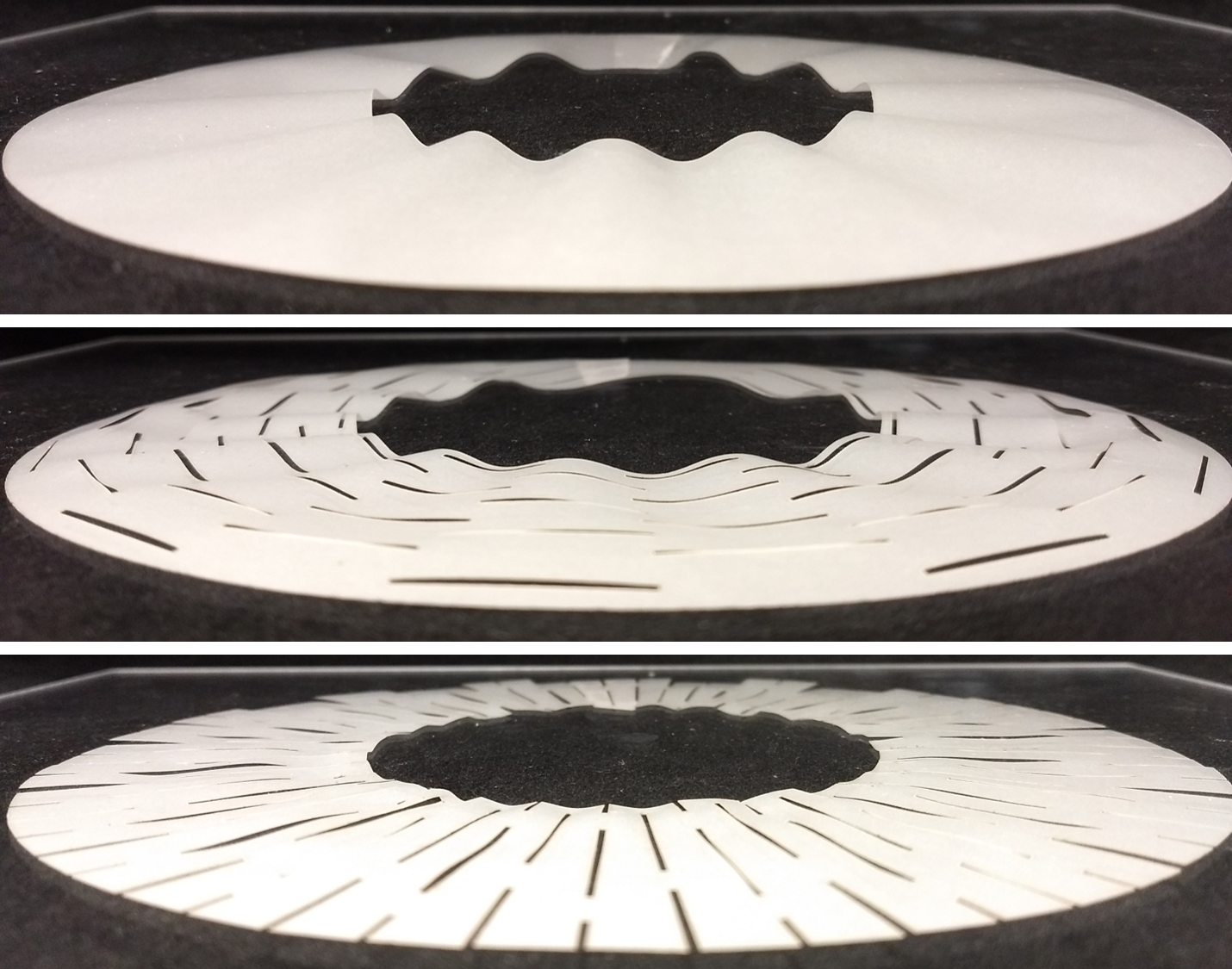}
	\caption{(a) A conical annulus flattened under a piece of acrylic, with a small gap allowing for wrinkle formation. (b) Azimuthal slits do not affect the pattern of wrinkles. (c) Radial slits result in azimuthal quadrupoles, minimizing interaction energy with the curvature monopole. When flattened, radial slits lead to a soft response with no wrinkles.}
	\label{fig:ConicalKirigami}
\end{figure}

Apart from the above global characterizations of frame mechanics, we also probe local measures such as the nonuniform displacement field over the entire frame, thereby allowing for a stronger test of the theory. Using grid lines etched into the paper, painted black to improve the contrast in imaging, we measure the displacement field of the frame by comparing its projected mesh just past buckling to a reference undeformed mesh. As the uniaxial tensile load prescribes the orientation of the induced quadrupoles, with just the scalar charge magnitudes as fitting parameters, we find the entire spatial deformation field is well captured within our image charge framework (red lines in Fig.~\ref{fig:defmesh}).

The quantitative success of our theory in describing the mechanics of isolated frames encourages us to take a step further and exploit the method of charges to analyze kirigami patterns, which now involves interactions between the charges in different holes. The elastic interaction energy of two planar quadrupoles $\b{Q}_1,\b{Q}_2$ a distance $\b{r}$ apart is given by \cite{Moshe2015PRE}
\begin{equation}
	E_{\mathrm{int}}=\dfrac{YQ_1Q_2}{\pi r^2}\cos(2\psi_1+2\psi_2)\ ,
\end{equation}
where the quadrupole angles $\psi_1,\psi_2$ are taken with respect to the pair separation $\b{r}$. To demonstrate that interacting elastic charges can fruitfully guide design of kirigami meta-materials, we shall focus on the simple problem of a flattened cone as an example of the inverse problem in kirigami mechanics. A conical frustum (either with an angle deficit or excess) when confined with a small gap in the plane is stressed due to its intrinsic geometry, a state that can be relaxed for a sufficiently thin sheet by wrinkling (Fig.~\ref{fig:ConicalKirigami}a). Patterning an appropriate kirigami design affords the sheet a new mechanism of locally relaxing in-plane stress \emph{without} wrinkling. For the regular circular cone, minimizing the energy of an interacting pair of quadrupoles with the background stress field of a positive disclination, we find (see SI) that the equilibrium configuration favours azimuthally aligned quadrupoles. Unlike squares that lock the quadrupole orientation to their diagonals, slits only permit quadrupolar charges perpendicular to their long axis. Hence, while azimuthal slits leave the wrinkles unaltered (Fig.~\ref{fig:ConicalKirigami}b), radial slits in a staggered array (which minimizes the charge-charge interactions) around the cone locally relax stress when flattened (Fig.~\ref{fig:ConicalKirigami}c). Similar slit patterns also relax stresses in a flattened e-cone \cite{muller2008conical} as shown in the SI.

In summary, we have proposed a useful elastic charge framework to understand kirigami mechanics in thin sheets with perforations.
By relating the challenging nonlinear problem of post-buckling mechanics to the simpler pre-buckling computation within the planar problem, we were able to quantitatively test the analytical predictions against experimental measurements through both global and local measures of deformation. The inclusion of interactions between charges also suggests our framework can advise possible design strategies to pattern kirigami meta-materials that permit engineering pathways to locally relax elastic stresses. Addressing nonlinear and thermal effects are promising directions for future work.

We thank James Pikul, Marc Miskin, Winston Lee, for their valuable insights and help with guiding the initial experiments. We thank Paul McEuen, Kyle Dorsey, Tanner Pearson, and Zeb Rocklin for very useful conversations throughout the project.
Work by IC was supported by a grant from the NSF DMREF program under grant DMR-1435829. Work by MJB was supported by the KITP grant PHY-1125915 and the NSF DMREF program, via grant DMREF-1435794. Work by DRN was primarily supported through the NSF DMREF program, via grant DMREF-1435999, as well as in part through the Harvard Materials Research and Engineering Center, via NSF grant DMR-1420570. MM acknowledges the USIEF Fulbright program. MM, SS \& MJB thank the Syracuse Soft \& Living Matter Program for support and the KITP for hospitality during completion of some of this work.


\pagebreak
\widetext
\begin{center}
\textbf{\large Kirigami mechanics as stress relief by elastic charges\\~\\
SUPPLEMENTARY INFORMATION\\}
	\vspace{1.3em}
	Michael Moshe$^{a,b}$, Edward Esposito$^{c}$, Suraj Shankar$^{b,d}$ Baris\\
	Bircan$^e$, Itai Cohen$^c$, David R. Nelson$^a$, and Mark J. Bowick$^{b,d}$\\
\textit{$^a$Department of Physics, Harvard University, Cambridge, Massachusetts 02138, USA.\\
$^b$Physics Department and Syracuse Soft and Living Matter Program,\\ Syracuse University, Syracuse, NY 13244, USA.\\
$^c$Laboratory of Atomic and Solid State Physics, Cornell University, Ithaca, NY 14853, USA.\\
$^d$Kavli Institute for Theoretical Physics, University of California, Santa Barbara, CA 93106, USA.\\
$^e$School of Applied and Engineering Physics, Cornell University, Ithaca, NY 14853, USA.}
\end{center}
\setcounter{equation}{0}
\setcounter{figure}{0}
\setcounter{table}{0}
\setcounter{page}{1}
\makeatletter
\renewcommand{\theequation}{S\arabic{equation}}
\renewcommand{\thefigure}{S\arabic{figure}}
\renewcommand{\bibnumfmt}[1]{[S#1]}
\renewcommand{\citenumfont}[1]{S#1}
\thispagestyle{empty}
\section{I.\hspace{1em} Experimental Methods}
The square frames are prepared by cutting them from full sheets of parchment paper (with Young’s moduli varying from $8\times10^8$~Pa to $16\times10^8$~Pa) using a $40$~W Epilog Zing $CO_2$ laser. Vector cutting is performed at $10$\% power and $100$\% speed. For imaging the deformation, mesh grids are etched onto the specimens using the laser cutter's raster setting at $10$\% power and $60$\% speed, and the back surface of the square is painted with a black oil paint to improve the contrast between the mesh and the body of the square. The cutting pattern also includes long, narrow tabs at opposite corners of the square, with dimensions $3$~mm$\times15$~mm. The tabs hold the frame fixed on the device.

A custom built force measurement setup is used to obtain the force-extension data of the square frame samples. The setup consists of a translation stage, equipped with a micrometer screw gauge, mounted to a Haydon-Kerk rail guided linear actuator (RGS08KR-M57-0250-18). The translation stage faces a polished aluminum plate mounted to a Loadstar force measurement sensor ($3$~kg RSP1 load cell with DI-1000U interface). The linear rail system is interfaced with Windows using the proprietary IDEA Drive Software and driven by a stepper IDEA Drive Kit (PCM4826E-K). The load cell is interfaced with MATLAB and force measurements are collected using a custom MATLAB script. At each displacement, the load cell takes $200$ measurements over $8$ seconds which are then averaged to produce the single measurement value reported. Our protocol uses $20$ such measurements in steps of $50$~$\mu$m, followed by measurements in steps of $500$~$\mu$m until the frame tears.

For each trial, the tab at one corner of the specimen is fixed with a fast setting 3M Super Glue (Ethyl cyanoacrylate) to the aluminum plate, and the tab at the opposite corner is glued to the surface of the translation stage. After a trial, the specimen and glue are cut away from the surfaces using a razor. Remaining residues are removed from the surfaces using BSI UN-CURE debonder. The surfaces are then allowed to dry for several minutes between trials.

The zero of the load cell must be recalibrated between trials. With the specimen glued to the apparatus, it is difficult to determine by eye the proper zero of the displacement, and thereby to set the force to zero. Therefore, we begin each trial with the specimen under slight, but visible, compressive buckling. During extension the samples will shift from this compressed state to a stretched state. By aligning the transition region for the data on samples with different thicknesses and aspect ratios we obtain the scaling plots shown in the manuscript.

The cones are cut from the same Glama Natural parchment paper, of intermediate thickness ($0.015$~cm). A full annulus is cut first. Then a $15^{\circ}$ wedge is either added or removed for the regular or the excess cone respectively, and the annulus is glued back together. The cones are photographed in their full three-dimensional configurations, and then are flattened under a quarter-inch thick piece of acrylic and photographed again in the flattened state.

\begin{figure}
	\includegraphics[width=0.9\textwidth]{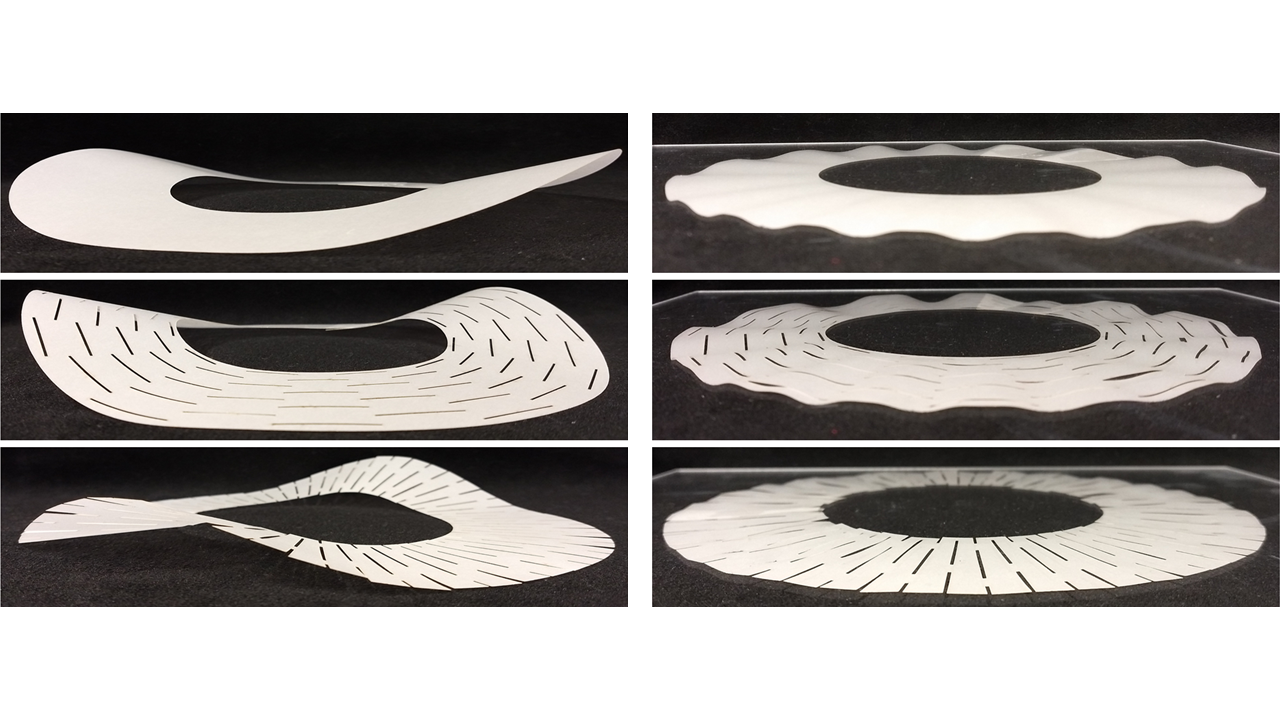}
	\centering
	\caption{A conical annulus with an angle surplus (e-cone). In the three images on the left, the e-cone (with and without the kirigami slit pattern) achieves its equilibrium shape in $3d$. On the right, the same annulus is flattened under an acrylic plate. Both the intact non-perforated e-cone as well as the one with azimuthal slits develop wrinkles when flattened, while the radial kirigami pattern of slits relaxes stresses in the flattened e-cone (as in the case of the regular cone shown in the main text), leaving it devoid of wrinkles.
	}
	\label{fig:econe}
\end{figure}

\section{II.\hspace{1em} Interacting quadrupoles on a flattened kirigami cone}
In the case of both an angle deficit (regular cone) or an excess (e-cone), the planar state of the cone is stressed and the Airy stress function is $\chi=(Y\Delta\phi/8\pi)r^2\ln(r/R)$ \cite{SISeung88}, where $\Delta\phi$ is the angle excess or deficit from $2\pi$, $Y$ the $2d$ Young's modulus and $R$ is the size of the sample. Both radial $\sigma_{rr}$ and hoop $\sigma_{\phi\phi}$ stresses are non-vanishing, while $\sigma_{r\phi}=0$. For not too big samples with a sufficiently large core excised out, these stresses are relaxed in the absence of kirigami patterns by the formation of radial wrinkles when the sample is flattened in the plane, as seen for the regular cone in the main text and for e-cones here in Fig.~\ref{fig:econe}.

Considering a periodic kirigami pattern of slits, as the quadrupole-quadrupole interaction decays quadratically with their separation \cite{SIMoshe2015PRE}, we restrict our analysis to nearest neighbour quadrupole interactions, which is the dominant term. Hence we have a pair of quadrupoles $\b{Q}_1$ and $\b{Q}_2$ at a separation $\b{r}_{12}$ in the presence of a background stress. Aligning the local coordinate frame so that the $x$-axis conincides with the azimuthal direction, we have the total elastic energy
\begin{equation}
	E=\dfrac{Y}{\pi r_{12}^2}Q_1Q_2\cos(2\psi_1+2\psi_2-4\theta_{12})-\dfrac{Q_1}{4}\sigma_0\cos2\psi_1-\dfrac{Q_2}{4}\sigma_0\cos2\psi_2\ ,
\end{equation}
	where $\sigma_0=\sigma_{rr}-\sigma_{\phi\phi}=\Delta\phi Y/(4\pi)$, $\theta_{12}$ is the angle made by $\b{r}_{12}$ with the $x$-axis and $\psi_1,\psi_2$ are the angles of the quadrupoles measured from the $x$-axis. This energy is minimized when the quardupoles align with the external field ($\psi_1=\psi_2=0$) and the angle of their pair separation is $\theta_{12}=\pi/4$. So a staggered array of quadrupoles aligned in the azimuthal direction relaxes the stressed state of a flattened cone. The geometry of a slit is such that when pulled, negative partial disclinations form at its ends and the positive partial disclinations are spread around its length leading to the induced quadrupole direction being essentially orthogonal to the long axis of the slit. Using this, we immediately find that a radial staggered array of slits will locally relax stresses in a flattened cone.

\end{document}